

\input harvmac
\def\rhob{{\rho\kern-0.465em \rho}}

\def\ontopss#1#2#3#4{\raise#4ex \hbox{#1}\mkern-#3mu {#2}}

\setbox\strutbox=\hbox{\vrule height12pt depth5pt width0pt}

\def\strut{\relax\ifmmode\copy\strutbox\else\unhcopy\strutbox\fi}
\nref\rslat{L. Slater, Proc. Lond. Math. Soc. (2) 54 (1952) 147.}
\nref\rgol{H. G{\"o}llnitz, {\it Einfache Partitionen} (unpublished),
Diplomarbeit W.S. 1960, G{\'o}ttingen, 65 pp; and
 J. Reine Angew. Math. 225 (1967) 154.}
\nref\rgor{B. Gordon, Duke. Math. J. 32 (1965) 741.}
\nref\randc{G.E. Andrews, Proc. Am. Math. Soc. 18 (1967) 945.}
\nref\randa{G.E. Andrews, in {\it The theory and applications of
special functions}, ed. R. Askey (Academic Press, New York, 1975)}
\nref\randb{G.E. Andrews,{\it Encyclopedia of Mathematics and its
Applications},vol. 2, Addison-Wesley, pp. 113-116.}

\nref\rbresa{D.M. Bressoud, Quart. J. Math. Oxford (2) 31 (1980) 385.}
\nref\rbresb{D.M. Bressoud, Lecture Notes in Math. 1395 (1987) 140.}
\nref\rbresc{D.M. Bressoud, in {it Ramanujan Revisited}, pp. 57-67,
G.E. Andrews et al eds., Academic Press, 1988.}
\nref\rbosform{B.L. Feigin and D.B. Fuchs, Funct. Anal. Appl. 16 (1982) 114;
P. Goddard, A. Kent and D. Olive, Comm. Math. Phys. 103 (1986) 105;
A. Meurman and A. Rocha-Caridi, Comm. Math. Phys. 107 (1986) 263.}
\nref\rfermi{R. Kedem, T.R. Klassen, B.M. McCoy and E. Melzer,
Phys. Letts. B 307 (1993) 68.}
\nref\rmel{E. Melzer, Supersymmetric Analogues of the Gordon-Andrews
Identities and Related TBA Systems, hepth 9412154.}
\nref\rber{A. Berkovich, Nucl. Phys. B4341 (1994) 315.}
\nref\rbmo{A. Berkovich, B.M. McCoy, and W. Orrick, {\it Polynomial
Identities, Indices, and Duality for the $N=1$ Superconformal Model
$SM(2,4\nu)$}, hepth 9507072}
\nref\rts{M. Takahashi and M. Suzuki, Prog. of Theo. Phys 48 (1972) 2187.}
\nref\rab{G.E. Andrews and R.J. Baxter, J. Stat. Phys. 47 (1987) 297.}
\nref\raab{A.K. Agarwal, G.E. Andrews and D.M. Bressoud, J. Indian Math. Soc.
51 (1987) 57.}
\nref\rfq{O. Foda and Y-H. Quano, Virasoro character Identities from the
Andrews-Bailey Construction, hepth 9408086.}
\nref\rbm{A. Berkovich and B.M. McCoy, Letts.in Math. Phys. (in press)
hepth 9507072.}
\nref\rfb{P.J. Forrester and R.J. Baxter, J. Stat. Phys. 38 (1985) 435.}
\nref\rwar{S. O. Warnaar, {\it Fermionic solution of the
Andrews-Baxter-Forrester model II: proof of Melzer's polynomial
identities},hepth 9508079.}
\nref\randd{G.E.  Andrews, Cont. Math. 16 (1994) 141.}
\nref\rcv{S. Cecotti and C. Vafa, Comm. Math. Phys. 157 (1993) 139.}
\nref\rcfiv{S. Cecotti, P. Fendley, K. Intriligator and C. Vafa,
Nucl. Phys. B 386 (1992) 405.}

  \Title{\vbox{\baselineskip12pt\hbox{BONN-TH-95-15}
  \hbox{ITPSB 95-29}
  \hbox{hep-th/9508110}}}
  {\vbox{\centerline{Generalizations of the Andrews-Bressoud
Identities}
\centerline{for the $N=1$ Superconformal Model $SM(2,4\nu)$ }}}
  \centerline{Alexander Berkovich~\foot{berkovic@axpib.physik.uni-bonn.de}}

  \bigskip\centerline{\it Physikalisches Institut der}
  \centerline{\it Rheinischen Friedrich-Wilhelms Universit{\"a}t Bonn}
  \centerline{\it Nussallee 12}
  \centerline{\it D-53115 Bonn, Germany}
  \bigskip
  \centerline{and}
  \bigskip
  \centerline{ Barry~M.~McCoy~\foot{mccoy@max.physics.sunysb.edu}}

  \bigskip\centerline{\it Institute for Theoretical Physics}
  \centerline{\it State University of New York}
  \centerline{\it Stony Brook,  NY 11794-3840}
  \bigskip

  \Date{\hfill 8/95}

  \eject

\centerline{\bf Abstract}
 We present generalized Rogers-Ramanujan identities which relate the
fermi and bose forms of all the characters of the superconformal model
$SM(2,4\nu).$ In particular we show that to each bosonic form of the character
there is an infinite family of distinct fermionic $q-$ series
representations.

\newsec{Introduction}
Generalized Rogers-Ramanujan identities are identities between
$q-$series which are part of a very rich area
of mathematics that includes:

1. $q-$series;~~~~~~~~~~~~~~~~~~~~~~~~~~~~~~~~~~~~2. Partition identities;

3. Path space counting problems;~~~~~~~ 4. Modular functions;

5. Polynomial identities;~~~~~~~~~~~~~~~~~~~~6. Conformal field theory;

7. Basic Hypergeometric Series;~~~~~~~~~~~8. Continued Fractions;

9. Dilogarithm identities;

10. Characters and branching functions of Kac-Moody algebras;

11. Solvable models of statistical mechanics;

12. Thermodynamic Bethe's Ansatz;

13. Holonomic Differential equations.

It is almost a ``folk theorem'' that
there is a $1-1$ isomorphism
between these concepts. This multiplicity of connections is both a
blessing in the number of insights provided to mathematics and physics
in the last 30 years and a curse in  that the explanation of any new
result needs to be made  in many different forms if it is to be
intelligible to all those  interested in the subject.
This general dilemma of presentation is well illustrated by the historical
background of the subject of this talk.

The history begins with the generalized Rogers-Ramanujan
identities that relate sums to products which are items 34 and 36
in the justly famous 1952 list of Lucy Slater~\rslat
\eqn\slata{\sum_{k=0}^{\infty}{(-q^{{1\over 2}})_k q^{{k(k+2)\over 2}}
\over (q)_k}=
{(-q^{{1\over 2}})_{\infty}\over (q)_{\infty}}\prod_{k=1}^{\infty}
(1-q^{4k-{1\over 2}})(1-q^{4k-{7\over 2}})(1-q^{4k})}
and
\eqn\slatb{\sum_{k=0}^{\infty}{(-q^{{1\over 2}})_k q^{{k^2\over 2}}
\over (q)_k}=
{(-q^{{1\over 2}})_{\infty}\over (q)_{\infty}}\prod_{k=1}^{\infty}
(1-q^{4k-{3\over 2}})(1-q^{4k-{5\over 2}})(1-q^{4k})}
where we use the notation
\eqn\qsyb{(A)_k=\cases{\prod_{j=0}^{k-1}(1-Aq^j)& if $k=1,2,\cdots$\cr
1&$k=0.$\cr}}
These identities were given a partition theoretic interpretation by
G{\"o}llnitz~\rgol~ and Gordon~\rgor~and this interpretation was  soon
thereafter generalized by Andrews~\randc~ as the following theorem:

{\it
Let $\nu$ and a be integers with $0<a \leq \nu.$ Let $C_{\nu,a}(n)$ be
the number of partitions of $n$ into parts which are neither $\equiv
2~({\rm mod}~4)$ nor $\equiv 0,\pm(2a-1)({\rm mod }~4\nu).$ Let
$D_{\nu,a}(n)$ denote the number of partitions of $N$ of the form
$n=\sum_{i=1}^{\infty}if_i$ with $f_1+f_2\leq a-1$ and for all $1\leq
i$}
\eqn\cond{f_{2i-1}\leq 1~~{\rm and}~f_{2i}+f_{2i+1}+f_{2i+2}\leq \nu
-1.}
{\it Then $C_{\nu,a}=D_{\nu,a}(n).$}

The corresponding generalization of the analytic result of
Slater
{}~\slatb~ was only given by
Andrews in 1975~\randa-~\randb~and the full generalization of
Slater's results was
finally given by
Bressoud~\rbresa--\rbresb~ who proved that for $s=1,3,5,\cdots,2\nu-1$
\eqn\andbres{\eqalign{&\sum_{n_2,\cdots,n_{\nu}=0}^{\infty}\
{(-q^{{1\over 2}})_{N_2}q^{{1\over 2}N_2^2+N_3^2+\cdots +N_{\nu}^2+\
N_{(s+3)/2}+N_{(s+5)/2}+\cdots N_{\nu}}\over (q)_{n_2}(q)_{n_3}\cdots\
 (q)_{n_{\nu}}}\cr
&={(-q^{{1\over 2}})_{\infty}\over (q)_{\infty}}\prod_{k=1}^{\infty}\
(1-q^{2\nu k})(1-q^{2\nu k-{s\over 2}})(1-q^{2\nu k -{4\nu -s\over
2}})\cr}}
where
\eqn\ndefin{N_k=\sum^{\nu}_{j=k} n_j}
The passage from ~\slata--\slatb~ to~ \andbres~ goes through partition
identities, path identities, continued fractions, basic
hypergeometric series and modular functions.

However  many more extensions of these results are possible
and the key to these extensions is the identification of
the identities~\andbres~ with the $N=1$ superconformal
field theory $SM(2,4\nu).$ This connection is made by using Jacobi's
triple product identity
\eqn\tprod{\sum_{j=0}^{\infty}z^j
q^{j^2}=\prod_{j=0}^{\infty}(1-q^{2j+2})(1+z q^{2j+1})(1+z^{-1}
q^{2j+1}),}
to rewrite the product side of \andbres~ as
\eqn\bose{{(-q^{{1\over 2}})_{\infty}\over
(q)_{\infty}}\sum_{j=-\infty}^{\infty}\
(q^{j(8\nu j +4\nu -2s)/2}-q^{(2j +1)(4\nu j +s)/2})}
and using the identity
\eqn\ident{(z)_N=\sum_{j=0}^{N}{N \atopwithdelims[] j}_q z^j
q^{j(j-1)/2},}
where (for latter use) we  use the slightly unconventional definition of the
$q$-binomial coefficient
\eqn\qbin{{l\atopwithdelims[] m}=\cases{(q)_l\over (q)_m (q)_{l-m}&if
$0\leq m \leq l$\cr
1&if $m=0,l\leq -1$\cr
0&otherwise,\cr}}
to rewrite the sum side of~\andbres~as
\eqn\fsum{\sum_{m_1,n_2,\cdots,n_{\nu}\geq 0}^{\infty}{q^{{m^2_1\over
2}-m_1 N_2 +\sum_{j=2}^{\nu}N_{j}^2+\sum_{j={s+3\over 2}}^{\nu}N_j}\over
(q)_{n_2} \cdots(q)_{n_{\nu}}}{N_2 \atopwithdelims[] m_1}_q.}
These expressions are to be compared with the bosonic form of
the characters of the
general $N=1$ superconformal model $SM(p,p')$~
{}~\rbosform
\eqn\schi{{\hat\chi}^{(p,p')}_{r,s}(q)={\hat\chi}_{p-r,p'-s}^{(p,p')}(q)=
{(-q^{\epsilon_{r-s}})_{\infty}\over
(q)_{\infty}}\sum_{j=-\infty}^{\infty}\bigl( q^{{j(jpp'+rp'-sp)\over 2}}
-q^{{(jp+r)(jp'+s)\over 2}}\bigr)}
where
\eqn\epsden{\epsilon_a=\cases{{1\over2}&if $a$ is even (Neveu-Schwarz
(NS) sector)\cr 1& if $a$ is odd (Ramond (R) sector)\cr}}
Here $r=1,2,\cdots,p-1$ and $s=1,2,\cdots,p'-1$ and $p$ and ${(p'-p)\over 2}$
are coprime.
Thus we see that
{}~\andbres~are the
characters of the Neveu-Schwarz sector of the $N=1$ superconformal
field theory $SM(2,4\nu)$. The sum expression~\fsum~is seen to be in
the canonical form of a fermionic representation of conformal field
theory characters~\rfermi.

The question now naturally arises of what corresponds to the
Andrews-Bressoud identities~\andbres~in Ramond sector of the field
theory and a partial answer to this was given by Melzer~\rmel~who found
generalizations of~\fsum~for Ramond characters with $s=2$ and $2\nu.$

Recently the authors, in collaboration with W. Orrick, have used the
methods which were developed to prove polynomial and character identities
for the minimal models $M(p,p+1)$~\rber~to
extend these results to all values of $s$ in the Ramond
sector~\rbmo. Indeed, it was found that there was not one but two
generalizations of~\fsum. But even more can be done and subsequent
investigation has revealed that in both the Neveu-Schwarz and
Ramond sectors there are an infinite number of fermionic forms for the
characters. The goal of this note is to present these results and to
indicate their proof. In fact, we will go further and extend
the character identities to the families of polynomial identities.
Some of these polynomial identities were presented for the first time
in ~\rbmo~ and some of them are new.

The plan of this note is as follows. In sec. 2 we will present both
the recent results of~\rbmo~and our subsequent generalizations. The
character identities are given in (2.33)-(2.35) and (2.37), (2.40).
The most general polynomial identities are given in (2.42), (2.43).
In sec. 3 we will indicate how the new results can be obtained from the
formalism of~\rbmo. In sec. 4 we conclude with a few comments about
the future direction of this research.

\newsec{Polynomial and Character Identities}

In this section we will separately define the polynomials which
generalize the fermionic and bosonic forms of the characters introduced in
sec. 1. We will then present the relations between these two forms
which generalize the Andrews-Bressoud identities~\andbres.

\subsec{Fermionic Polynomials}

In ~\rbmo~ the set of variables  $({\vec m},{\vec n})$ were introduced
which satisfy the relations
\eqn\countg{\eqalign{n_1+m_1&={1\over 2}(L+m_1-m_2)-a_1\cr
n_2+m_2&={1\over 2}(L+m_1+m_3)-a_2\cr
n_i+m_i&={1\over 2}(m_{i-1}+m_{i+1})-a_i,~~~~{\rm for}~3\leq i \leq \nu-1\cr
n_{\nu}+m_{\nu}&={1\over 2}(m_{\nu-1}+m_{\nu})-a_{\nu}\cr}}
where $n_i$ and $m_i$ are integers and the components $a_i$  of the vector
${\vec a}$ are either integers or half integers.
This system is closely related to the $TBA$ equations for the $XXZ$-model
((3.9) of~\rts) with anisotropy
\eqn\anis{\gamma=\pi{(2\nu-1)\over 4\nu}.}

In terms of these variables we then introduce the
following polynomials to generalize the
Fermi form of the characters~\fsum
\eqn\fpoly{F^{(\nu,n)}_{r',s'}(L,q)=\sum_{{\cal D}_{r',s'}} q^{Qf+Lf_{n,s'}}
\prod_{i=1}^{\nu}{n_i+m_i\atopwithdelims[] n_i}_q}
where the quadratic form $Qf$ and linear form $Lf_{n,s'}$ are
\eqn\qf{Qf={m_1^2\over 2}-m_1 N_2+\sum_{j=2}^{\nu} N^2_j
{}~~~{\rm and}~~~Lf_{n,s'}=n{m_1\over 2}+\sum_{l=\nu-s'+1}^{\nu} N_l,}
the ranges of $r'$ and $s'$ are
\eqn\ranges{r'=0,1,2,\cdots,\nu-2~~{\rm and}~~s'=0,1,2,\cdots,\nu-1,}
the relation between $s'$ and $s$ is
\eqn\ssp{s=\cases{2\nu-2s'&if $n$ is odd\cr
2\nu-2s'-1&if $n$ is even,\cr}}
and  the vector ${\vec a}$ is
defined by
\eqn\indefn{\eqalign{{\vec a}&={\vec a}^{(r')}+{\vec a}^{(s')}\cr
a^{(k)}_i&=\cases{{1\over 2}
(\delta_{i,{\nu}}-\delta_{i,\nu-k})&for $0\leq k \leq \nu-2$\cr
{1\over 2}(\delta_{i,\nu}+\delta_{i,1}) &for $k=\nu-1$.\cr}}}

The domain of summation, ${\cal D}_{r',s'}$ is best described in terms of
${\vec n}$ and $m_{\nu}$ which are subject to the constraint derived
from~\countg~
\eqn\constraint{L=(n_1+a_1)+m_{\nu}+\sum_{i=2}^{\nu}(2i-3)(n_i+a_i).}
All other variables are given by
\eqn\eqnva{m_1=N_2-n_1}
\eqn\eqnvb{m_i=m_{\nu}+2\sum_{j=i+1}^{\nu} (j-i)(n_j+a_j),
{}~~~~~i=2,3,\cdots,\nu-1.}
Keeping in mind that (from the definition ~\qbin)
\eqn\qbinnewdef{{{\rm neg.~ int.}\atopwithdelims[] 0}_q=1,}
we define ${\cal D}_{r',s'}$ for $s'\geq r'$ as the union of the sets
of solutions to~\constraint~satisfying
\eqn\extra{\eqalign{0:~~~&n_i,~m_{\nu}\geq 0,\cr
1:~~~&n_{\nu}=0,m_{\nu}=-2,~n_1,\cdots, n_{\nu-1}\geq 0,\cr
2:~~~&n_{\nu}=n_{\nu-1}=0,~m_{\nu}=-4,~n_1,\cdots n_{\nu-2}\geq 0,\cr
&~~~~\cdots\cr
r':~~~&n_{\nu}=n_{\nu-1}=\cdots =
n_{\nu-r'+1}=0,~~~m_{\nu}=-2r',~~~n_1,\cdots,n_{\nu-r'}\geq 0;\cr}}
and for $s'<r'$ the definition is the same  as above with $r'\rightarrow s'$.

Using the asymptotic formula
\eqn\qbinlim{\lim_{A\rightarrow\infty}{A\atopwithdelims[]B}_q={1\over(q)_B}}
and the simple consequence of~\countg
\eqn\ancons{n_i+m_i=L+m_1+n_i-2\sum_{j=2}^i (j-1)(n_j+a_j)-
2\sum_{j=i+1}^{\nu} (i-1)(n_j+a_j);~~~i\geq 2}
along with~\eqnva, we find
\eqn\pcf{\lim_{L\rightarrow\infty}F_{r',s'}^{(\nu,n)}(L,q)=F^{(\nu,n)}_{s'}(q)}
which holds for all $r'$ where
\eqn\fdefn{F^{(\nu,n)}_{s'}(q)=\sum_{m_1,n_2,\cdots,n_{\nu}\geq 0}
{q^{Qf+Lf_{n,s'}}\over(q)_{n_2}(q)_{n_3}\cdots (q)_{n_\nu}}{N_2
\atopwithdelims[]m_1}_q.}

\subsec{Bosonic Polynomials}

To generalize the bosonic character formula to a polynomial expression
we introduce the $q-$trinomial coefficients $T_{n}(L,A;q^{{1\over 2}})$
of Andrews and Baxter~\rab~ defined in terms of
\eqn\appa{{L,A-n;q\atopwithdelims()A}_2=\sum_{j \geq 0}t_n(L,A;j),~~~~~n\in Z}
with
\eqn\appaa{t_n(L,A;j)={q^{j(j+A-n)}(q)_L\over (q)_j (q)_{j+A}(q)_{L-2j-A}}}
as
\eqn\tndf{T_n(L,A;q^{{1\over 2}})=q^{{L(L-n)-A(A-n)\over 2}}
{L,A-n;q^{-1}\atopwithdelims() A}_2.}
These trinomial coefficients satisfy the recursion relations proven in~\randd~
and~\rbmo
\eqn\triident{\eqalign{T_n(L,A;q^{{1\over 2}})&=T_n(L-1,A+1;q^{{1\over 2}})+
T_n(L-1,A-1;q^{{1\over 2}})\cr
&+q^{L-{n+1\over 2}}T_n(L-1,A;q^{{1\over 2}})+(q^{L-1}-1)T_n(L-2,A;
q^{{1\over 2}}).\cr}}
We note the elementary property
\eqn\prop{T_n(L,A;q^{{1\over 2}})=T_n(L,-A;q^{{1\over 2}})}
and remark that
\eqn\tnsym{T_n(L,A;-q^{{1\over 2}})=\cases{(-1)^{L+A}T_n(L,A;q^{{1\over 2}})
&for $n$ even\cr
T_n(L,A;q^{{1\over 2}})& for $n$ odd.\cr}}
Consequently, $T_n(L,A;q^{{1\over 2}})$ is actually a polynomial in $q$ for
$n$ odd or for $n$ even and $L+A$ even, while for $n$ even
and $L+A$ odd $T_n(L,A,q^{{1\over 2}})$ contains only odd powers of
$q^{{1\over 2}}.$

We then have the following definition of bosonic polynomials:

1) For the Neveu-Schwarz sector with $n=0$
\eqn\bosepolyns{\eqalign{B^{(\nu,0)}_{r',s'}(L,q)=&
\sum_{j=-\infty}^{\infty}(-1)^j q^{\nu
j^2+(s'+{1\over 2})j}\biggl(T_0(L,2\nu j +s'-r';q^{{1\over 2}})\cr
&+T_0(L,2\nu j +s'+1+r';q^{{1\over 2}})\biggr);}}

2) For the  Ramond sector with $n=-1$
\eqn\bosepolyr{B^{(\nu,-1)}_{r',s'}(L,q)=\sum_{j=-\infty}^{\infty}
(-1)^j q^{\nu j^2+s'j}\sum_{i=-r'}^{r'}
(-1)^{r'+i}T_1(L,2\nu j+s'+ i;q^{{1\over 2}}).}
For other $n\geq 1$ we define $B^{(\nu,n)}_{r',s'}$ from the recursion
relation
\eqn\brecrel{B^{(\nu,n+2)}_{r',s'}(L,q)=B^{(\nu,n)}_{r',s'}(L+1,q)+
q^{-(n+1)/2}B^{(\nu,n)}_{r',s'}(L,q).}
Then using the relation ,which is derivable from~\triident,
\eqn\tnrecrel{\eqalign{&T_{-n}(L+1,A;q^{{1\over 2}})
+q^{-(n+1)/2}T_{-n}(L,A;q^{{1\over 2}})\cr
&=T_{-(n+2)}(L,A+1;q^{{1\over 2}})+T_{-(n+2)}(L,A-1;q^{{1\over 2}})\cr
&~~~~+(q^{(n+1)/2}+
q^{-(n+1)/2})T_{-(n+2)}(L,A;q^{{1\over 2}})\cr}}
we find, for example, for $n=1$ in the Ramond sector
\eqn\bosepolyrp{\eqalign{&B^{(\nu,1)}_{r',s'}(L,q)=\cr
&\sum_{j=-\infty}^{\infty}(-1)^j q^{\nu
j^2+s'j}\biggl(T_{-1}(L,2\nu j +s'-r';q^{{1\over 2}})+
T_{-1}(L,2\nu j +s'+1+r';q^{{1\over 2}})\cr
&~~~~~~~~~~~~~~~+T_{-1}(L,2\nu j +s'-r'-1;q^{{1\over 2}})+\
T_{-1}(L,2\nu j +s'+r';q^{{1\over 2}})\biggr);\cr}}
and for $n=2$ in the Neveu-Schwarz sector
\eqn\btwo{\eqalign{&B^{(\nu,2)}_{r',s'}(L,q)=\cr
&\sum_{j=-\infty}^{\infty}(-1)^j q^{\nu j^2+(s'+{1\over 2})j}\
\biggl(T_{-2}(L,2\nu j+s'-r'-1;q^{{1\over 2}})+T_{-2}(L, 2\nu j+s'-r'+1;
q^{{1\over 2}})\cr
&~~~~~T_{-2}(L,2\nu j+s'+r'+2;q^{{1\over 2}})+T_{-2}(l,2\nu j+s'+r';q^{{1\over
2}})\cr
&~~~~~(q^{{1\over 2}}+q^{-{1\over 2}})[T_{-2}(L,2\nu j+s'-r';q^{{1\over 2}})+
T_{-2}(L,2\nu j+s'+r'+1;q^{{1\over 2}})]\biggr).}}

In general, $B^{(\nu,n)}_{r',s'}(L,q)$ polynomials are given in terms of
$T_{-n}-$trinomials.

Using the limiting formula
\eqn\tzlim{\lim_{L\rightarrow \infty}T_n(L,A;
q^{{1\over 2}})=\cases{{(-q^{{(1-n)\over 2}})_{\infty}+
(q^{{(1-n)\over 2}})_{\infty}\over 2(q)_{\infty}}&if $L-A$ is even\cr
{(-q^{{(1-n)\over 2}})_{\infty}-
(q^{{(1-n)\over 2}})_{\infty}\over 2(q)_{\infty}}& if $L-A$ is odd.\cr}}
and noting the special case
\eqn\tolim{\lim_{L\rightarrow \infty}
T_{1}(L,A;q^{{1\over 2}})={(-q)_{\infty}\over (q)_{\infty}}}
we find the relation between the limit
\eqn\blim{B^{(\nu,n)}_{s'}(q)=\lim_{L\rightarrow \infty}
B^{(\nu,n)}_{r',s'}(L,q)}
and the characters~\schi
\eqn\pcb{\eqalign{B^{(\nu,0)}_{s'}(q)&=
{\hat \chi}^{(2,4\nu)}_{1,2\nu -2s'-1}(q)\cr
B_{s'}^{(\nu,1)}(q)&=
B_{s'}^{(\nu,-1)}(q)
={\hat \chi}^{(2,4\nu)}_{1,2\nu -2s'}(q)\cr}}
which holds for all $r'.$

\subsec{Bose/Fermi  Identities}

In~\rbmo~we found the following relation between the fermi and bose
forms of the characters for $n=0,\pm1:$

1) For the Neveu-Schwarz sector with $n=0$ we have (the original
Andrews-Bressoud identities~\andbres)
\eqn\bfns{B_{s'}^{(\nu,0)}(q)=F_{s'}^{(\nu,0)}(q);}

2) For the Ramond sector with $n=1$
\eqn\bfrp{B_{s'}^{(\nu,1)}(q)=\cases{F^{(\nu,1)}_{s'}(q)+
F^{(\nu,1)}_{s'-1}(q)& for $ s'\neq 0 $\cr
2F^{(\nu,1)}_{0}(q)& for $s'=0;$\cr}}

3) For the Ramond sector with $n=-1$
\eqn\bfrm{F_{s'}^{(\nu,-1)}(q)=\cases{B_{\nu-1}^{(\nu,-1)}(q)&for
$s'=\nu-1$\cr
B_{s'}^{(\nu,-1)}(q)+B_{s'+1}^{(\nu,-1)}(q)&for $s'\neq\nu-1$\cr}}
or, equivalently
\eqn\bfrmt{B_{s'}^{(\nu,-1)}(q)=\sum_{l=s'}^{\nu-1}
(-1)^{l+s'}F_{l}^{(\nu,-1)}(q).}

These results can be extended to all integers $n$ as follows:

Neveu-Schwarz:

\eqn\nsneg{\eqalign{{\vec F}^{(\nu,-2n)}(q)&
={1\over (-q^{{1\over 2}})_n}\prod_{j=1}^n
K(\nu,-2j){\vec B}^{(\nu,0)}(q)~~~{\rm for}~~n\geq 1\cr
{\vec F}^{(\nu,2n)}(q)&=q^{-n^2/2}(-q^{{1\over 2}})_n
\prod_{j=0}^{n-1}K^{-1}(\nu,2j){\vec B}^{(\nu,0)}(q)~~{\rm for}~~n\geq 1\cr}}
where we use the vector notation
\eqn\notved{\eqalign{{\vec F}_{r'}^{(\nu,n)}(L,q)|_{s'}&=
F^{(\nu,n)}_{r',s'}(L,q)~~~~
{\vec B}^{(\nu,n)}_{r'}(L,q)|_{s'}=B^{(\nu,n)}_{r',s'}(L,s)~~{\rm and}\cr
{\vec F}^{(\nu,n)}(q)|_{s'}&=F^{(\nu,n)}_{s'}(q)~~~~
{\vec B}^{(\nu,n)}(q)|_{s'}=B_{s'}^{(\nu,n)}(q)\cr}}
and the matrix $K(\nu,n)$ is defined for $0\leq i,j \leq \nu-1$ by
\eqn\kmat{[K(\nu,n)]_{i,j}=\delta_{i,j}(\delta_{i,0}+\delta_{i,\nu-1}+
q^{(n+1)/2}+q^{-(n+1)/2})+\delta_{i,j+1}+\delta_{i,j-1}}

Ramond:

\eqn\ramneg{\eqalign{{\vec F}^{(\nu,-1-2n)}(q)&
={1\over (-q)_n}\prod_{j=1}^n K(\nu,-1-2j)
C {\vec B}^{(\nu,-1)}(q)~~{\rm for}~~n\geq 1\cr
{\vec F}^{(\nu,2n+1)}(q)&=
(-1)_{n+1}q^{-n(n+1)/2}\prod_{j=0}^nK^{-1}(\nu,2j-1)
C{\vec B}^{(\nu,-1)}(q)~~{\rm for}~~n\geq 0\cr}}
where for $0\leq i,j\leq \nu-1$
\eqn\cdefn{[C]_{i,j}=\delta_{i,j}+\delta_{i+1,j}.}

For $n\geq -1$ these character identities can be extended to
polynomial identities. Defining $B^{(\nu,n)}_{r',s'}(L,q)$ through
{}~\brecrel~we have
\eqn\nspoly{\eqalign{{\vec F}^{(\nu,2n)}_{r'}(L,q)&=
\prod_{j=0}^{n-1}K^{-1}(\nu,2j){\vec B}^{(\nu,2n)}_{r'}(L,q),~~~~~~~n\geq 1\cr
{\vec F}^{(\nu,2n+1)}_{r'}(L,q)&=\prod_{i=0}^n
K^{-1}(\nu,2j-1)C{\vec B}^{(\nu,2n+1)}_{r'}(L,q),~~~~n\geq 0.\cr}}
along with the two identities proven in~\rbmo
\eqn\nspolysec{\eqalign{{\vec F}^{(\nu,0)}_{r'}(L,q)&=
{\vec B}^{(\nu,0)}_{r'}(L,q)\cr
{\vec F}^{(\nu,-1)}_{r'}(L,q)&=C{\vec B}^{(\nu,-1)}_{r'}(L,q).\cr}}

\newsec{Sketch of Proof}

 The proof of the results of the previous section starts with the
recursion relations established in~\rbmo~ for the fermionic polynomials
{}~\fpoly ~which holds for all allowed values of $s'$
\eqn\nseqn{\eqalign{h^{(n)}_0(L)&=
h^{(n)}_1(L-1)+(q^{L+{n-1\over2}}+1)h^{(n)}_0(L-1)+
(q^{L-1}-1)h^{(n)}_0(L-2),\cr
h^{(n)}_{r'}(L)&=h^{(n)}_{r'-1}(L-1)+h^{(n)}_{r'+1}(L-1)+
q^{L+{n-1\over 2}}h^{(n)}_{r'}(L-1)+
(q^{L-1}-1)h^{(n)}_{r'}(L-2)\cr
&~~~~~~~~~~~~~~~~~~~~~~~~~~~~~~~~~~~~~~~~~~~~~~{\rm for}~1\leq r'\leq \nu-3,\cr
h^{(n)}_{\nu-2}(L)&=h^{(n)}_{\nu-3}(L-1)+
q^{L+{n-1\over 2}}h^{(n)}_{\nu-2}(L-1)+
q^{L-1}h^{(n)}_{\nu-2}(L-2);\cr}}
where we
note that the first and the last equations follow from the middle equation
if one introduces $h^{(n)}_{-1}(L)$ and $h^{(n)}_{\nu-1}(L)$ satisfying
\eqn\hmo{h^{(n)}_{-1}(L)=h^{(n)}_{0}(L)}
and
\eqn\hnmo{h^{(n)}_{\nu-1}(L)=h^{(n)}_{\nu-2}(L-1).}
For $\nu=2$ there is only the single equation
\eqn\nutns{h^{(n)}_0(L)=(1+q^{L+{n-1\over 2}})h^{(n)}_0(L-1)+
q^{L-1}h^{(n)}_0(L-2).}
These equations specify the fermionic polynomials~\fpoly~uniquely if in
addition we impose the boundary conditions
\eqn\fbound{\eqalign{F_{r',s'}^{(\nu,n)}(0,q)&=\delta_{r',s'}\cr
F_{r',s'}^{(\nu,n)}(1,q)&=\delta_{s',r'+1}+\delta_{s',r'-1}+
\delta_{s',0}\delta_{r',0}+q^{n+1\over 2}\delta_{r',s'}\cr}}

It is important to observe that  any set of functions
$h_{r'}^{(n)}(L)$
defined by ~\nseqn~ are not independent but satisfy the relation
\eqn\hnrecrel{h_r^{(n+2)}(L)=h^{(n)}_{r'}(L+1)+
q^{-{n+1\over2}}h_{r'}^{(n)}(L).}
To prove this for the generic equation of ~\nseqn~
we add ~\nseqn~with $L\rightarrow L+1$ to ~\nseqn~
multiplied by $q^{-{n+1\over2}}$ and define $X^{(n)}_{r'}(L)=
h_{r'}^{(n)}(L+1)+q^{-{n+1\over 2}}h_{r'}^{(n)}(L)$ to find
\eqn\exn{\eqalign{X^{(n)}_{r'}(L)=X^{(n)}_{r'-1}(L-1)&+X^{(n)}_{r'+1}(L-1)
+q^{L+{n-1\over 2}}X^{(n)}_{r'}(L-1)+(q^{L-1}-1)X^{(n)}_{r'}(L-2)\cr
&{\rm for}~~1\leq r'\leq \nu-3}}
which is the generic equation  of~\nseqn~with $n\rightarrow n+2.$
The remaining equations of ~\nseqn~ are treated similarly. Thus the
relation~\hnrecrel~follows. In~\rbmo~we have demonstrated that the bosonic
polynomials $B^{(\nu,n)}_{r',s'}(L,q)$ with $n=0,\pm 1$ satisfy~\nseqn. The
relation~\hnrecrel~shows that it is true for all $n\geq -1$.

We now proceed to establish the recursion relation
\eqn\fident{{\vec F}^{(\nu,n)}_{r'}(L+1,q)+
q^{-(n+1)\over 2}{\vec F}^{(\nu,n)}_{r'}(L,q)=
K(\nu,n){\vec F}^{(\nu,n+2)}_{r'}(L,q)}
with $K(\nu,n)$ given by~\kmat. This is proven by first
using~\hnrecrel~as applied to the fermionic polynomials~\fpoly~to
show that the left hand side of~\fident~is expressed as a linear
combination of $F_{r',s'}^{(\nu,n+2)}(L,q)$ for $0\leq s' \leq
\nu-1.$ The matrix $K$ which specifies this linear combination is then
determined by demanding that~\fident~hold for $L=0$ and $1$ using
the boundary conditions~\fbound~along with~\nseqn.

We will now prove the results of sec. 2. First consider the
character identities~\nsneg~and~\ramneg. Let $L\rightarrow\infty$
in~\fident~to obtain the recursion relation on the fermionic sums~\fdefn
\eqn\charident{(1+q^{-(n+1)/2}){\vec F}^{(\nu,n)}(q)=
K(\nu,n){\vec F}^{(\nu,n+2)}(q).}
If $n$ is even this may be solved in terms of ${\vec F}^{(\nu,0)}$
as
\eqn\enenn{\eqalign{{\vec F}^{(\nu,-2m)}(q)&={1\over(-q^{{1\over 2}})_m}
\prod_{j=1}^{m}K(\nu,-2j){\vec F}^{(\nu,0)}(q)~~{\rm for}~~m\geq 1\cr
{\vec
F}^{(\nu,2m)}(q)&=q^{-m^2/2}(-q^{{1\over 2}})_m\prod_{j=0}^{m-1}
K^{-1}(\nu,2j){\vec F}^{(\nu,0)}(q)~~{\rm for}~~m\geq 1\cr}}
from which ~\nsneg~follows by using the bose/fermi identity~\bfns.

Similarly if $n$ is odd we solve~\charident~in terms of ${\vec
F}^{(\nu,-1)}(q)$ and use the identity~\bfrm~which was proven
in~\rbmo~to obtain~\ramneg.

The corresponding results for polynomials given by~\nspoly~is a simple
consequence of the fermionic identity~\fident, the recursive
definition~\brecrel~of the bosonic polynomials $B^{(\nu,n)}_{r',s'}(L,q)$ and
identities~\nspolysec.

\newsec{Discussion}

There are several features of the general theory of Rogers-Ramanujan
type identities which are very clearly seen in the results summarized in
sec. 2 and because of their importance deserve to be discussed in further
detail.

The first of these features is the occurrence in the range of summation
${\cal D}_{r',s'}$~\extra~of the fermionic sum~\fpoly~ of solutions of
the constraint ~\countg~where some of the variable are negative.
This is an entirely new aspect of fermionic sums which has first been
seen in this problem. It is, however, a very general feature, which,
once recognized, is seen to occur in many of the fermionic character
formulae of nonunitary models $M(p,p').$ This was explicitly seen in
{}~\rbmo~for the minimal model $M(2\nu-1,4\nu)$ whose character
polynomials are obtained from the polynomials of $SM(2,4\nu)$ by the
duality transformation $q\rightarrow {1\over q}.$ These extra terms are
needed to extend the fermionic polynomials of the minimal model
$M(p,p')$ from the special subset of characters considered in our
previous study~\rbm~to the general case.

The second feature of our results which deserves prominent discussion
is the fact that we have developed a rather complete study of the
phenomenon of linear combinations. Indeed, with  exceptions such as
the $n=0$ Neveu-Schwarz and the $n=1,~s=2\nu$ and the
$n=-1,~s=2$ Ramond characters all of our formulae for the bosonic
characters ${\hat \chi}^{(2,4\nu)}_{r,s}(q)$ involve linear combinations
of two or more fermionic series of the canonical form~\fpoly. Such
linear combinations of the canonical fermionic series have been seen
before~\rfermi,~\raab,~\rbresc~and~\rfq~in several special cases, but in this
present study of the $SM(2,4\nu)$ model it is clear that these linear
combinations are a mandatory integral part of the theory and that,
indeed, it was the failure to consider linear combinations which
prevented previous authors from finding a complete set of character formulae.

In our study~\rbm~ of the models $M(p,p')$ we explicitly restricted
ourselves to cases where these linear combinations did not occur and the
generalization of these results will certainly involve linear
combinations. As an example of the sort of new phenomena which takes place
we consider the dual transformation $q\rightarrow {1\over q}$ which sends
$SM(2,4\nu)$ into $M(2\nu-1,4\nu)$~\rbmo. For this transformation we
define
\eqn\dual{\eqalign{{\vec f}^{(\nu,n)}_{r'}(q)&=\lim_{L\rightarrow
\infty}q^{L(L+n)\over 2}{\vec F}^{(\nu,n)}_{r'}(L,{1\over q})\cr
{\vec b}_{r'}^{(\nu,n)}(q)&
=\lim_{L\rightarrow \infty}q^{L(L+n)\over 2}
{\vec B}^{(\nu,n)}_{r'}(L,{1\over q})\cr}}
and find from~\rbmo~
\eqn\minb{\eqalign{{\vec b}_{r'}^{(\nu,0)}(q)|_{s'}&=q^{(s'-r')^2\over
2}\chi^{(2\nu-1,4\nu)}_{\nu-r'-1,2\nu-2s'-1}(q)+
q^{(s'+r'+1)^2\over 2}\chi^{(2\nu-1,4\nu)}_{\nu+r',2\nu-2s'-1}(q)\cr
{\vec b}_{r'}^{(\nu,-1)}(q)|_{s'}&
=q^{(s'-r')(s'-r'-1)\over 2}\chi^{(2\nu-1,4\nu)}_{\nu-r'-1,2\nu-2s'}(q)+
q^{(s'+r')(s'+r'+1)\over 2}\chi^{(2\nu-1,4\nu)}_{\nu+r',2\nu-2s'}(q)\cr}}
where
\eqn\roca{\chi_{r,s}^{(p,p')}(q)=\chi_{p-r,p'-s}^{(p,p')}(q)=
{1\over (q)_{\infty}}\sum_{j=-\infty}^{\infty}(q^{j(jpp'+rp'-sp)}-
q^{(jp+r)(jp'+s)}).}
We also find the fermionic forms
\eqn\limdfermi{\eqalign{\lim_{L\rightarrow\infty}
q^{{L(L+n)\over 2}}F_{r',s'}^{(\nu,n)}(L,q^{-1})&=
\sum_{{\tilde{\bf m}}{\rm -restrictions}}
{q^{\Phi_n({\tilde{\bf m}},r',s')}\over
(q)_{{\tilde m}_1}(q)_{{\tilde m}_2}}\cr
&\times \prod_{i=3}^{\nu}{((1-{\bf B}){\tilde{\bf m}})_i-a_i^{(r')}-a_i^{(s')}
\atopwithdelims[]{\tilde m}_i}_q\cr}}
where
\eqn\mtilde{{\tilde{\bf m}}^t=(n_1,m_2,m_3,\cdots ,m_{\nu})}
\eqn\kunya{\Phi_n({\tilde{\bf m}},r',s')={1\over 2}{\tilde{\bf m}}{\bf B}
{\tilde{\bf m}}+L_n({\tilde{\bf m}},s')+C_n(r',s')}
\eqn\dulf{2L_n({\tilde{\bf m}},s')={\tilde m}_{\nu}-{\tilde m}_{\nu-s'}
+{\tilde m}_1\delta_{s',\nu-1}+(2{\tilde m}_1+{\tilde m}_2)
(n+\delta_{s',\nu-1})}
\eqn\duct{4C_n(r',s')=s'-r'+(1+2n)\delta_{s',\nu-1}}
the matrix
${\bf B}$ defined by its matrix elements
\eqn\bmat{({\bf B})_{j,k}=\cases{2&for $j=k=1$\cr
\delta_{k,2}&for $j=1,~2\leq k\leq\nu$\cr
\delta_{j,2}&for $k=1,~2\leq j\leq\nu$\cr
{1\over 2}\delta_{j,2}\delta_{k,2}+\delta_{j,k}-{1\over 2}\delta_{j,k+1}-
{1\over 2}\delta_{j,k-1}-{1\over 2}\delta_
{j,\nu}\delta_{k,\nu}&otherwise,}}
the inhomogeneous vectors $a_i^{(s')}$ and $a_i^{(r')}$ are given by
{}~\indefn~and the restrictions on the summation variables
${\tilde{\bf m}}$ are
\eqn\restrik{{\tilde m}_i-{\tilde m}_{\nu}=
\bigl({\vec v}^{(s')}+{\vec v}^{(r')}
\bigr)_{i-1}({\rm mod}~2);~~~~~ i=2,3,\cdots,\nu}
with
\eqn\vdef{({\vec v}^{(k)})_i=k\theta(1\leq i\leq\nu-k-1)+
(\nu-1-i)\theta(k>0)\theta(\nu-k-1<i\leq\nu-1)}
where $k=0,1,\cdots,\nu-1$.

Then we find from~\fident~ the dual analogue of~\charident~
\eqn\dualcharident{{\vec f}^{(\nu,n)}_{r'}(q)=
q^{n+1\over 2}K(\nu,n){\vec f}^{(\nu,n+2)}_{r'}(q)}
and thus find for $M(2\nu-1,4\nu)$ the results
\eqn\duaresultsa{\eqalign{{\vec f}^{(\nu,-2n)}_{r'}(q)&=
q^{-n^2/2}\prod_{j=1}^nK(\nu,-2j){\vec b}_{r'}^{(\nu,0)},~~~{\rm
for}~~n\geq 1\cr
{\vec f}^{(\nu,2n)}_{r'}(q)&
=q^{-n^2/2}\prod_{j=0}^{n-1}K^{-1}(\nu,2j){\vec
b}_{r'}^{(\nu,0)}~~~{\rm for}~~n\geq 1\cr}}
and
\eqn\dularesultsb{\eqalign{{\vec f}_{r'}^{(\nu.2n+1)}(q)&
=q^{-{n(n+1)\over 2}}\prod_{j=0}^n K^{-1}(\nu,2j-1)
C{\vec b}_{r'}^{(\nu,-1)}(q)~~{\rm for}~~n\geq 0\cr
{\vec f}^{(\nu,-1-2n)}_{r'}(q)&=q^{-{n(n+1)\over 2}}\prod_{j=1}^n
K(\nu,-1-2j)C{\vec b}_{r'}^{(\nu,-1)}~~{\rm for}~~n\geq 1\cr}}
where we have used (6.23)-(6.25) of ~\rbmo~ rewritten in the form
\eqn\bfdual{{\vec f}_{r'}^{(\nu,0)}(q)={\vec b}^{(\nu,0)}_{r'}(q)
{}~~{\rm and}~~{\vec f}^{(\nu,-1)}_{r'}(q)=C{\vec b}^{(\nu,-1)}_{r'}(q).}
We thus find that for the minimal model $M(2\nu-1,4\nu)$ there are an
infinite number of ways in which the characters
$\chi^{(2\nu-1,4\nu)}_{r,s}(q)$ are given in term of linear
combinations of fermionic $q-$series. It is to be expected that this is
a general feature of the theory of all of the models $M(p,p')$ and
thus there is a great deal of generalization of the results
of~\rbm~ which can be carried out.

The expression for the characters~\minb~ of the $M(2\nu-1,4\nu)$ in
terms of the bosonic polynomials $B_{r',s'}^{(\nu,0)}(L,{1\over q})$~
and $B_{r',s'}^{(\nu,-1)}(L,{1\over q})$~\bosepolyns-\bosepolyr~reveals yet
another important feature of Rogers-Ramanujan type identities; namely the
characters of $M(2\nu-1,4\nu)$ which have a well known polynomial
representations in terms of $q-$binomials~\rfb~ also have a (different)
representation in terms of $q-$trinomials. Further examples of
trinomial representations of characters of
minimal models  are given by Andrews and Baxter~\rab~ for the $M(2,7)$ model
and by Warnaar~\rwar~ for the unitary model $M(p,p+1).$
Just as $q-$ binomials are a
natural basis of functions for a spin ${1\over 2}$ $XXZ$ system so are the
$q-$trinomials the natural basis for a spin $1$ $XXZ$ system. Moreover there
are extensions of the $q-$trinomials to $q-$multinomials~\randd~which are
related to higher spins.
It is expected, but not yet presented in the literature, that
each order of multinomial will lead to a separate polynomial representation of
the characters~\roca~of $M(p,p').$ Moreover, it is likely that spin-$1$
$SM(2,4\nu)$ polynomials presented here also have higher spin analogs.

As our final comment of this discussion we return from the technical
extensions of Rogers-Ramanujan identities which have been suggested by
the details of the computations on the $SM(2,4\nu)$ model to the
``folk theorem'' stated in the introduction.

The assertion of the ``folk theorem'' is that  the 13 items on the list
are not independent subjects but all originate in a more basic
mathematical concept such that if properly formulated the intrinsic
relations of the items on the list would be manifest in a general
fashion which would obviate the need for detailed proofs of individual
connections. It is our belief that this principle is counting as
seen mathematically in combinatorics and  physically in the notion
of the statistics of excitations.

This fundamental importance of counting is embodied in the choice of
variables~\countg~and has been made very explicit both the
fermionic counting computations presented in ~\rber~and ~\rbmo. This
variable choice originates in the treatment of the thermodynamic
Bethe's Ansatz equations of the spin ${1\over 2}$ $XXZ$ chain of ref.~\rts. It
is this counting problem for a system with a finite number of fermionic
excitations parameterized by $L$ that unifies the first 12 items on the
list of sec. 1.

But the constraint equations~\countg~are obtained as a very special case
of the equations of~\rts~where the temperature $T$ is set precisely to
zero and $L$ is kept finite. The $q-$counting problems of~\rber~and
{}~\rbmo~are obtained from~\rts~in the limit $T\rightarrow 0,$
$L\rightarrow \infty$ with $TL$ fixed and in this limit all energy
momentum relations of the excitations are linearized. But in the complete
treatment of the
thermodynamic Bethe's Ansatz the limit $L\rightarrow \infty$ at a
fixed temperature is taken and the energy momentum relations are not
linearized. The result is a set of nonlinear
integral equations which explicitly involve the temperature $T.$
It would thus appear as if item 12 on the list is much more general
that the previous 11 items. The conclusion we draw from this is that
the full significance of the combinatorial basis of the thermodynamic
Bethe's Ansatz remains to be explored and that new Rogers-Ramanujan type
identities which would incorporate the physical concept of the nonlinear
dispersion relations are to be discovered.

Exactly what types of relations amongst TBA solutions should be
implied by the counting problem cannot of course be speculated on in
any detail. But surely it is of great importance that for the TBA
equations of a particular $N=2$ supersymmetric model it has been shown
that TBA equations can be solved in terms of the Painlev{\'e} III
differential equation~\rcv-\rcfiv. It is our belief that there is a counting
basis not only to the Painlev{\'e}
equations but to all holonomic systems of equations which are obtained
by a deformation procedure. It is this speculation which is implied by
the inclusion of the last item on the list of sec. 1.

\bigskip
\noindent
{\bf Acknowledgments}

We would like to thank the entire organizing committee for the
opportunity of participation in the conference ``Physique et Combinatoire''
and for the hospitality of the CIRM in the Facult{\' e} de Sciences de
Luminy. This work was partially supported by NSF grant DMR 9404747 and
by the Deutsche Forshungs Gemeinschaft.

\vfill
\eject
\listrefs

\vfill\eject

\bye
\end